\documentclass[review]{elsarticle}

\usepackage{amsmath}
\usepackage{amssymb}
\usepackage{enumitem}

\usepackage{color}

\newcommand{\TSYN}{\tilde I_{syn}}
\newcommand{\TSHD}{\tilde I_{shd}}
\newcommand{\TUNQ}{\tilde I_{unq}}
\newcommand{\eqdef}{=\mathrel{\mathop:}}

\usepackage{setspace} 
\usepackage{amssymb}
\usepackage{nameref}

\usepackage[top=3cm, bottom=2cm, left=4cm, right=4cm]{geometry}

\newcommand{\eq}[1]{equation~\ref{eq:#1}}

\newcommand{\appRef}[1]{Appendix~\ref{app:#1}}

\usepackage{graphicx}

\usepackage{lineno,hyperref}
\modulolinenumbers[5]

\begin{document}

\begin{frontmatter}

\title{Partial Information Decomposition as a Unified Approach to the Specification of Neural Goal Functions}

\author{Michael Wibral\fnref{MEGUnit}}
\fntext[MEGUnit]{wibral@em.uni-frankfurt.de, MEG Unit, Brain Imaging Center, Goethe University, Heinrich Hoffmann Stra{\ss}e 10, 60528 Frankfurt am Main, Germany}

\author{Viola Priesemann\fnref{Goettingen}}
\fntext[Goettingen]{Department of Non-linear Dynamics, Max Planck Institute for Dynamics and Self-Organization, \& Bernstein Center for Computational Neuroscience, G\"{o}ttingen, Germany} 

\author{Jim W.\,Kay\fnref{Glasgow}}
\fntext[Glasgow]{Department of Statistics,University of Glasgow, Glasgow, G12 8QQ, UK}

\author{Joseph T.\,Lizier\fnref{USyndney}}
\fntext[USyndney]{School of Civil Engineering, The University of Sydney, NSW, Australia}

\author{William A.\,Phillips\fnref{Stirling}}
\fntext[Stirling]{School of Natural Sciences, University of Stirling, Stirling, UK}


%
%

\begin{abstract}
In many neural systems anatomical motifs are present repeatedly, but despite their structural similarity they can serve very different tasks.
A prime example for such a motif is the canonical microcircuit of six-layered neo-cortex, which is repeated across cortical areas, and is involved in a number of different tasks (e.g.sensory, cognitive, or motor tasks). This observation has spawned interest in finding a common underlying principle, a 'goal function', of information processing implemented in this structure. By definition such a goal function, if universal, cannot be cast in processing-domain specific language (e.g. 'edge filtering', 'working memory'). Thus, to formulate such a principle, we have to use a domain-independent framework. Information theory offers such a framework. However, while the classical framework of information theory focuses on the relation between one input and one output (Shannon's mutual information), we argue that neural information processing crucially depends on the combination of \textit{multiple} inputs to create the output of a processor. To account for this, we use a very recent extension of Shannon Information theory, called partial information decomposition (PID). PID allows to quantify the information that several inputs provide individually (unique information), redundantly (shared information) or only jointly (synergistic information) about the output. First, we review the framework of PID. Then we apply it to reevaluate and analyze several earlier proposals of information theoretic neural goal functions (predictive coding, infomax and coherent infomax, efficient coding). We find that PID allows to compare these goal functions in a common framework, and also provides a versatile approach to design new goal functions from first principles. Building on this, we design and analyze a novel goal function, called 'coding with synergy', which builds on combining external input and prior knowledge in a synergistic manner. We suggest that this novel goal function may be highly useful in neural information processing.


\end{abstract}


\end{frontmatter}


\section{Introduction}
In many neural systems anatomical and physiological motifs are present repeatedly in the service of a variety of different functions. A prime example is the canonical cortical microcircuit that is found in many different regions of the six-layered mammalian neocortex. These different regions serve various sensory, cognitive, and motor functions, but how can a common circuit be used for such a variety of different purposes? This issue has spawned interest in finding a common abstract framework within which the relevant information processing functions can be specified.

Several solutions for such an abstract framework have been proposed previously, among them approaches that still use semantics to a certain extent (predictive coding with its initial focus on sensory perception), teleological ones that prescribe a goal based on statistical physics of the organism and its environment (free energy principle) and information theoretic ones that focus on local operations on information (Coherent Infomax). While these are all encouraging developments, they also beg the question of how to compare these approaches, and how many more possibilities of defining new approaches of this kind exist. Ideally,  an abstract framework that would comprise these approaches as specific cases would be desirable. This article suggests a possible starting point for the development of such a unifying framework.

By definition this framework cannot be cast in processing-domain specific language, such as `edge-filtering' or `face perception’, or `visual working memory’, for example, but must avoid any use of semantics beyond describing the elementary operations that information processing is composed of \footnote{To be truly generic, the framework should also avoid to resort too strongly to semantics in terms of ``survival of the organism'' as even that maybe not desirable for each and every individual organism in certain species. This is because ``programmed death'' will allow a more rapid turnover of generations and thereby more rapid evolutionary adaptation.}. A framework that has these properties is information theory. In fact, information theory is often criticized exactly for its lack of semantics, i.e. for ignoring the \emph{meaning} of the information that is processed in a system. As we will demonstrate here, this apparent shortcoming can be a strength when trying to provide a unified description of the goals of neural information processing. Moreover, by identifying separate component processes of information processing, information theory provides a meta-semantics that serves to better understand what neural systems do at an abstract level (for more details see \cite{wibral_bits_2015}). Last, information theory is based on evaluating probabilities of events and thereby closely related to the concepts and hypotheses of probabilistic inference that are at the heart of predictive coding theory \cite{hohwy_predictive_2013,clark_whatever_2013,lee_hierarchical_2003,rao_predictive_1999}. Thus information theory is naturally linked to the domain-general semantics of this and related theories.

Based on the domain-generality of information theory several variants of information theoretic goal functions for neural networks have been proposed. The optimization of these abstract goal functions on artificial neural networks leads to the emergence of properties also found in biological neural systems -- this can be considered an amazing success of the information theoretic approach given that we still know very little about general cortical algorithms. This success raises hopes for finding unifying principles in the flood of phenomena discovered in experimental neuroscience. Examples of successful, information-theoretically defined goal functions are Linsker's infomax \cite{Linsker1988} -- producing receptive fields and orientation columns similar to those observed in primary visual cortex V1  \cite{Bell_independent_1997}, recurrent infomax -- producing neural avalanches, and an organization to synfire-chain like behaviour \cite{Tanaka_recurrent_2009}, and coherent infomax \cite{phillips_discovery_1995}. The goal function of coherent infomax is to find coherent information between two streams of inputs from different sources, one conceptualized as sensory input, the other as internal contextual information. As coherent infomax requires the precomputation of an integrated receptive field input as well as an integrated contextual input to be computable efficiently (and thereby, in a biologically plausible way), the theory predicted the recent discovery of two distinct sites of neural integration in neocortical pyramidal cells \cite{larkum_cellular_2013}. For details see the contribution of Phillips to this special issue. We will revisit some of these goal functions below and demonstrate how they fit in the larger abstract framework aiming at a unified description that is presented here.

Apart from the desire for a unified description of the common goals of repeated anatomical motifs, there is a second argument in favor of using an abstract framework. This argument is based on the fact that a large part of neural communication relies on axonal transmission of action potentials and on their transformation into post-synaptic potentials by the receiving synapse. Thus, for neurons, there is only one currency of information. This fact has been convincingly demonstrated by the successful rewiring of sensory organs to alternative cortical areas that gave rise to functioning, sense-specific  perception (see for example the cross-wiring, cross-modal training experiments in \cite{melchner_visual_2000}). In sum, neurons only see the semantics inherent in the train of incoming action potentials, not the semantics imposed by the experimenter. Therefore, a neurocentric framework describing information processing must be necessarily abstract. From this perspective information theory is again a natural choice.

Classic Shannon information theory, however, mostly deals with the transmission of information through a communication channel with one input and one output variable. In a neural setting this would amount to asking how much information present at the soma of one cell reaches the soma of another cell across the connecting axons, synapses and dendrites, or how much information is passed from one circuit to another. Information processing, however, comprises more operations on information than just its transfer. A long tradition dating back all the way to Turing has identified the elementary operations of information as information transfer, active storage, and modification. Correspondingly, measures of information transfer have been extended to cover more complex cases than Shannon's channels, incorporating directed and dynamic couplings \cite{Schreiber2000} and multivariate interactions \cite{Lizier2008}, and also measures of active information storage have been introduced \cite{Lizier2012a}. Information modification, seemingly comprising of subfunctions such as 
\emph{de novo} creation and fusion of information, however, has been difficult to define \cite{Lizier2013InfoMod}.
 
One reason for extending our view of information processing to more complicated cases is that even the most simple function from Boolean logic that any other logic function can be composed of (NAND, see for example \cite{Jaynes_Book}, chapter 1) uses two distinct input variables and one output. While such a logic function could be described as a channel between the two inputs and the outputs, this does not do justice to the way the two inputs interact with each other. What is needed instead is an extension of classic information theory to three way systems, describing how much information in the output of this Boolean function, or any other three-way processor of information, comes uniquely from one input, uniquely from the other input, how much they share about the output, and how much output information can only be obtained from evaluating both inputs jointly.

These questions can be answered using an extension of information theory called partial information decomposition (PID) \cite{PaulWilliamsPID,harder_bivariate_2013,Bertschinger2014,Griffith2014a}.

This article will introduce PID and show how to use it to specify a generic goal function for neural information processing. This generic goal function can then be adapted to represent previously defined neural information processing goals such as infomax, coherent infomax and predictive coding. This representation of previous neural goal functions in just one generic framework is highly useful to understand their differences and commonalities. Apart from a reevaluation of existing neural goal functions, the generic neural goal function introduced here also serves to define novel goals not investigated before.

The remainder of the text will first introduce partial information decomposition, and then demonstrate its use to decompose the total output information of a neural processor. From this decomposition we derive a generic neural goal function ``$G$'', and then express existing neural goal functions as specific parameterizations of G. We will then discuss how the use of $G$ simplifies the comparison of these previous goal functions and how it helps to develop new ones.

\section{Partial Information decomposition}
\label{sec:PID}

\begin{figure}
  \includegraphics[width=1\textwidth]{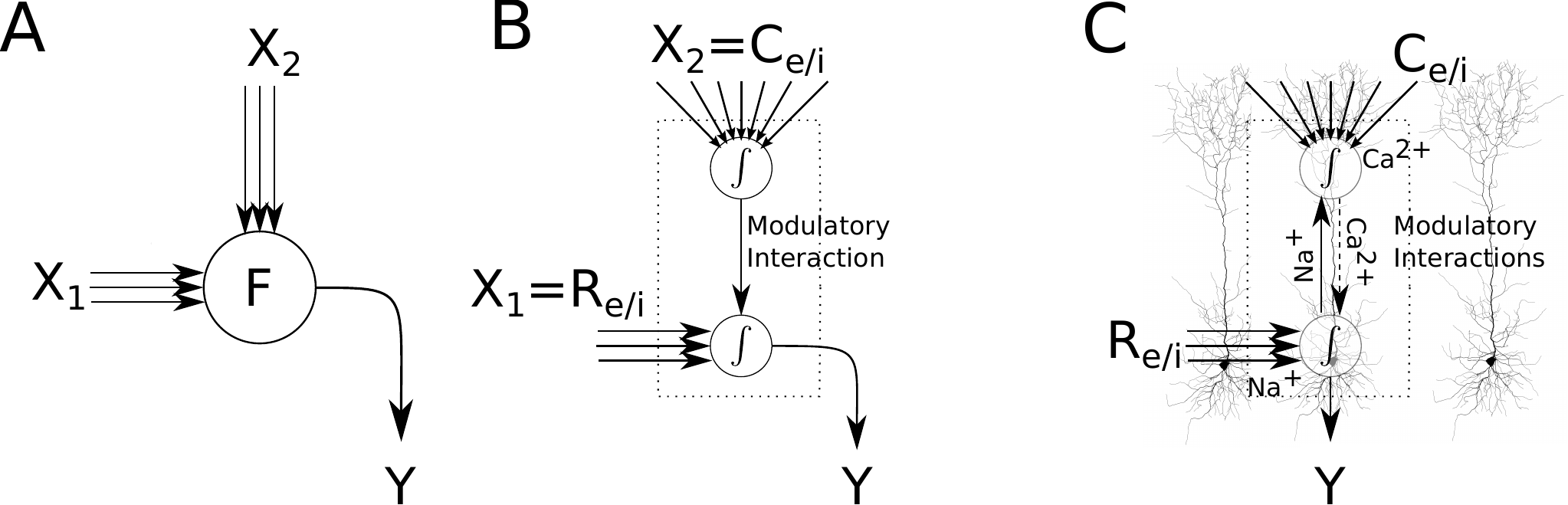}
\caption{Neural processors:(A) neural processor with multidimensional inputs $X_1$, $X_2$, and output $Y$. (B) Processor with local weighted summation of inputs as used in coherent infomax and in this study. To establish the link to the coherent infomax literature we identify the input $X_1$ with the receptive field input $R$, which may be excitatory (e) or inhibitory (i), and which is summed. In the same way, $X_2$ is identified with the contextual input $C$. (C) Overlay of the coherent infomax neural processor on a layer 5 pyramidal cells, highlighting potential parallels to existing physiological mechanisms. Layer 5 cells created with the TREES toolbox \cite{cuntz_rule_2010}, courtesy of Hermann Cuntz.}
\label{fig:NeuralProcessor}
\end{figure} 

In this section we will describe the framework of partial information decomposition (PID) to the extent that is necessary to understand the decomposition of the mutual information between the output $Y$ of a neural processor and a set of two inputs $X_1$, $X_2$ (Figure \ref{fig:NeuralProcessor}). The inputs themselves may be multivariate random variables but we will not attempt to decompose their contributions further. This is linked to the fact that in many neurons contextual and driving inputs are first summed separately before being brought to interact to produce the output. This summation strongly reduces the parameter space and thereby makes learning tractable -- see \cite{KayBookchapter1999,Kay2011}.\footnote{
Furthermore, the formulation of measures providing generally-accepted decompositions \cite{Bertschinger2014,Griffith2014a} at the present time are only defined for two variables \cite{Rauh2014}.}
Therefore, we limit ourselves to the PID of the mutual information between one ``left hand side'' or ``output'' variable $Y$ and two ``right hand side'' or ``input'' variables $X_1$, $X_2$. That is, we decompose the mutual information $I(Y: X_1,X_2)$ \footnote{As the concepts of unique, shared and synergistic information require a more fine grained distinction of how individual variables are grouped, we employ the following extended notation that was introduced in \cite{Bertschinger2014} and defined in \appRef{Notation}: ``:'' separates \emph{sets} of variables between which mutual information or partial information terms are computed, ``;'' separates multiple \emph{sets} of variables on one side of a partial information term, whereas ``,'' separates variables within a set that are considered jointly (see the \appRef{Notation} for examples).}, the total amount of information held in the set $\{ X_1, X_2 \}$ about $Y$:\footnote{See notational definitions in \appRef{Notation}.}
\begin{align}
 I(Y : X_1,X_2) & = \sum_{x_1 \in \mathcal{A}_{X_1}, x_2 \in \mathcal{A}_{X_2}, y \in \mathcal{A}_{Y}}{ p(x_1,x_2,y) \log_2{\frac{p(y|x_1,x_2)}{p(y)} } } \\
 	& = H(Y) - H(Y | X_1,X_2)~,
\end{align}
\noindent where the $\mathcal{A}_\cdot$ signifiy the support of the random variables and $H(\cdot)$, $H(\cdot | \cdot)$ are the entropy and the conditional entropy, respectively (see \cite{CoverBook} for definitions of these information theoretic measures).

The PID of this mutual information addresses the questions:

\begin{enumerate} 
\item What information does one of the variables, say $X_1$, hold individually about $Y$ that we can not obtain from any other variable ($X_2$ in our case)? This information is the \emph{unique information} of $X_1$ about $Y$: ~$I_{unq}(Y:X_1 \setminus X_2)$.
\item What information does the joint input variable $\{X_1; X_2\}$ have about $Y$ that we cannot get from observing both variables $X_1$, $X_2$ separately? This information is called the \emph{synergy}, or \emph{complementary information}, of $\{X_1;X_2\}$ with respect to $Y$: ~$I_{syn}(Y : X_1;X_2)$.
\item What information does one of the variables, again say $X_1$, have about $Y$ that we could also obtain by looking at  the other variable ($X_2$) alone? This information is the \emph{shared} information\footnote{Also known as \emph{redundant} information in \cite{PaulWilliamsPID}.} of $X_1$ and $X_2$ about $Y$: ~$I_{shd}(Y: X_1 ; X_2)$. 
\end{enumerate}

\begin{figure}[thpb]
\begin{center}
\includegraphics[width=0.5\textwidth]{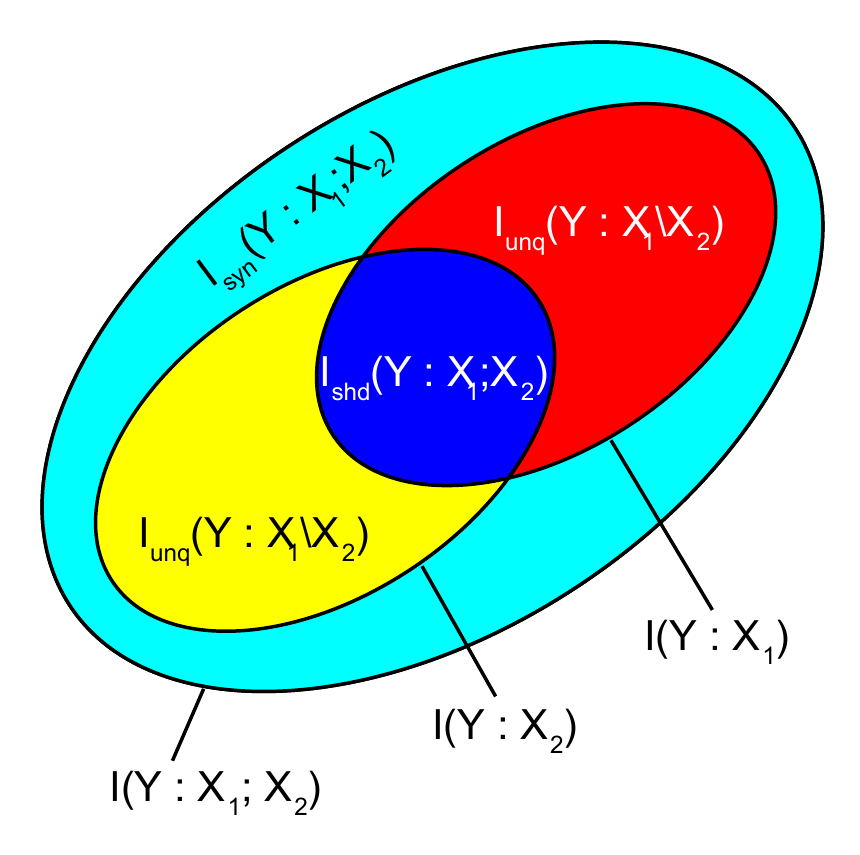}
\end{center}
\caption{Partial information diagram with both classical information terms (solid lines) and PID terms (color patches).}
\label{fig:PID} 
\end{figure}

\noindent Following \cite{PaulWilliamsPID}, the above three types of partial information terms together by definition provide all the information that the set $\{X_1,X_2\}$ has about $Y$, and other sources agree on this \cite{PaulWilliamsPID,Griffith2014a,harder_bivariate_2013,Bertschinger2014}, i.e.:

\begin{equation}
\label{eq:totalMI}
 I(Y:X_1,X_2) = I_{unq}(Y:X_1 \setminus X_2) + I_{unq}(Y:X_2 \setminus X_1) + I_{shd}(Y: X_1 ; X_2) + I_{syn}(Y: X_1 ; X_2)~,
\end{equation}
\noindent Figure \ref{fig:PID} is a graphical depiction of this notion by means of the partial information (PI-) diagrams introduced in \cite{PaulWilliamsPID}. In addition, there is agreement that the information one input variable has about the output should decompose into a unique and a shared part as:
\begin{equation}
\label{eq:singleMI}
\begin{split}
 I(Y:X_1) = I_{unq}(Y:X_1 \setminus X_2) + I_{shd}(Y: X_1 ; X_2)\\
 I(Y:X_2)=I_{unq}(Y:X_2\setminus X_1)+I_{shd}(Y:X_1;X_2) ~.
 \end{split}
\end{equation}

For the treatment of neural goal functions we have to furthermore give PID representations of the relevant conditional mutual information terms. These can be obtained from equations \ref{eq:totalMI} and \ref{eq:singleMI} as :
\begin{equation}
\label{eq:condMI}
\begin{split}
I(Y:X_1|X_2)= I_{unq}(Y:X_1\setminus X_2) + I_{syn}(Y:X_1;X_2)\\
I(Y:X_2|X_1)=I_{unq}(Y:X_2\setminus X_1) + I_{syn}(Y:X_1;X_2)~.
\end{split}
\end{equation}

\noindent Moreover, all parts of the PI-diagram are typically required to be positive to allow an interpretation as information terms.

Due to the pioneering work of Williams and Beer \cite{PaulWilliamsPID} it is now well established that neither unique, nor shared, nor synergistic information can be obtained from the definitions of entropy, mutual information and conditional mutual information in classical information theory.
Essentially, this is because we have an underdetermined system, i.e. we have fewer independent equations relating the output and inputs in classical information theory (three for two input variables) than we have PID terms (four for two input variables).
For at least one of these PID terms a new, axiomatic definition is necessary, from which the others then follow, as per equations \ref{eq:totalMI}-\ref{eq:condMI}. To date, the equivalent axiom systems introduced by Bertschinger and colleagues \cite{Bertschinger2014}, and by Griffiths and Koch \cite{Griffith2014a} have found the widest acceptance. They also yield results that are very close to an earlier proposal by Harder and colleagues \cite{harder_bivariate_2013}. All of these axiom systems lead to measures that are sufficiently close to a common sense view of unique, shared and synergistic information, and all satisfy equations \ref{eq:totalMI}-\ref{eq:condMI}. Hence, their exact details do not matter at first reading for the purposes of this paper, and will therefore be presented in \appRef{PID}.

The one exception to this statement is that we have to mention here already that shared information may arise in the frameworks of Bertschinger at al.\, \cite{Bertschinger2014}, Griffiths et al.\,\cite{Griffith2014a} , and also Harder et al.\, \cite{harder_bivariate_2013} for two reasons. First, there can be shared information because the two inputs $X_1$, $X_2$ have mutual information between them (termed \emph{source redundancy} in \cite{harder_bivariate_2013}, and source shared information here) -- this is quite intuitive for most. Second, shared information can arise because of certain \emph{mechanisms} creating the output  $Y$ (\emph{mechanistic redundancy} in \cite{harder_bivariate_2013}, mechanistic shared information here). This second possibility of creating shared information is less intuitive but nevertheless arises in all of the frameworks mentioned above. For example, the binary AND operation on two \emph{independent} (identically distributed) binary random variables creates $0.311$ bits of shared information in \cite{Bertschinger2014,harder_bivariate_2013,Griffith2014a}, and $0.5$ bits of synergistic mutual information, while there is no unique information about the inputs in its output.

\section{A generic decomposition of the output information of a neural processor}

We use PID in this section to decompose the information $H(Y)$ that is contained in the output of a general neural processor (Figure \ref{fig:NeuralProcessor}) with two input (sets) $X_1$ and $X_2$ and an output $Y$:
\begin{align}
\label{eq:Bandwidth}
H(Y) =& I(Y:X_1,X_2) + H(Y|X_1,X_2)\\
\nonumber =& I_{unq}(Y:X_1 \setminus X_2) + I_{unq}(Y:X_2 \setminus X_1) \\
\nonumber &+ I_{shd}(Y: X_1;X_2) + I_{syn}(Y: X_1;X_2) \\
\nonumber &+ H(Y|X_1,X_2) ~.
\end{align}

To arrive at a neural goal function we can add weight coefficients to each of the terms in the entropy decomposition above to specify how 'desirable' each one of one of these should be for the neural processor, i.e. we can specify a neural goal function $G$ as a function of these coefficients. Since all the terms in equation \ref{eq:Bandwidth} are non-overlapping, and the coefficients can be be chosen independently, this is the most generic way possible to specify such a goal function:
\begin{align}
\label{eq:GoalFunction}
G =& \Gamma_0 I_{unq}(Y:X_1 \setminus X_2) + \Gamma_1 I_{unq}(Y:X_2 \setminus X_1) \\
\nonumber &+ \Gamma_2 I_{shd}(Y: X_1;X_2) + \Gamma_3 I_{syn}(Y: X_1;X_2) \\
\nonumber &+ \Gamma_4 H(Y|X_1,X_2) ~.
\end{align}
which can also be rewritten with another set of of coefficients $\gamma_i$ as:
\begin{align}
\label{eq:GoalFunction_HY}
G =& \gamma_0 I_{unq}(Y:X_1 \setminus X_2) + \gamma_1 I_{unq}(Y:X_2 \setminus X_1) \\
\nonumber &+ \gamma_2 I_{shd}(Y: X_1;X_2) + \gamma_3 I_{syn}(Y: X_1;X_2) \\
\nonumber &+ \gamma_4 H(Y) ~,
\end{align}
\noindent  using $\gamma_i = \Gamma_i-\Gamma_4$ ($i=0\ldots3$), $\gamma_4 = \Gamma_4$ (and \eq{Bandwidth}).

Note that training a neural processor will obviously change the value of the goal function in equation \ref{eq:GoalFunction}, but of course also change the relative composition of the entropy in equation \ref{eq:Bandwidth}.

This decomposition of the entropy and its parametrization are closely modeled on the approach taken by Kay and Phillips in their formulation of another versatile information theoretic goal function (``$F$'', see below) for the coherent infomax principle \cite{phillips_discovery_1995,kay_contextually_1998,KayBookchapter1999,Kay2011}.

In general, we will choose the formulation used in equation \ref{eq:GoalFunction} because the conditional entropy does not overlap with the parts in the PI-diagram (Figure \ref{fig:PID}), but note that the formulation used in equation \ref{eq:GoalFunction_HY} may be useful when goals with respect to total bandwidth, rather than unused bandwidth, are to be made explicit. This could for example happen when neuronal plasticity acts to increase to total bandwidth of a neural processor\footnote{Along similar lines one may wish to add the total information in both inputs $H(X_1)$ and $H(X_2)$ (or,$H(X_1|Y,X_2)$ and $H(X_2|Y,X_1)$, respectively) to $G$. However, since the neural processor has only control over the output $Y$, changing the amount of this initial information in the inputs is beyond the scope of its goal function. }.

In the next sections we introduce coherent infomax and analyze it by means of PID. We then show how to (re-)formulate infomax, and predictive coding using specific choices of parameters for $G$. Last, we will introduce a neural goal function, called \emph{coding with synergy}, that explicitly exploits synergy for information processing.

\section{The coherent infomax principle and its goal function as seen by PID}

\subsection{The coherent infomax principle}
The coherent infomax principle (CIP) proposes an information theoretically defined neural goal function in the spirit of domain-independence laid out in the introduction, and a neural processor implementing this goal function \cite{phillips_discovery_1995,kay_contextually_1998,KayBookchapter1999,Kay2011}. The neural processor operates on information it receives from two distinct types of inputs $X_1$, $X_2$ and send the results to a single output $Y$ (see Figure \ref{fig:NeuralProcessor}). The two distinct types of input in CIP were described as driving and modulatory, formally defined by their distinct roles in local processing as detailed in the coherent infomax principles \ref{1}-\ref{4}, below. Here we will denote the driving input by $X_1$, and the contextual input by $X_2$.

In the mammalian brain the driving input $X_1$ includes, but is not limited to, both external information received from the sensors and information retrieved from memory. The contextual input $X_2$ arises from diverse sources as lateral long-range input from the same or different brain regions, descending inputs from hierarchically higher regions, and input via non-specific thalamic areas. Phillips, Clark and Silverstein \cite{phillips_functions_2015} provide a recent in-depth review of this issue in relation to the evidence for such distinct inputs from several disciplines.

The coherent infomax principle (CIP) states the following four goals of information processing:
\begin{enumerate}[label=\textbf{CIP.\arabic*}]
\item \label{1}The output $Y$ should transmit information that is shared between the two inputs, so as to enable the processor to preferentially transmit information from the driving inputs ($X_1$) that is supported by context-carrying information from internal sources elsewhere in the system arriving at input $X_2$. This is what the term 'coherent' refers to.
\item  \label{2} The output $Y$ could transmit \emph{some} information that is only in the driving input $X_1$, but not in the context, so as to enable that local processors transmit some information that is not related to the information currently available to it from elsewhere in the system.
\item \label{3} The output $Y$ should minimize transmission of information that is only in the contextual input $X_2$. This is necessary to ensure that the effects of the context do not become confounded with the effects of the drive and thereby reduce the reliability of coding .
\item \label{4} The output $Y$ should be optimally used in terms of bandwidth.
\end{enumerate}

To state these goals more formally, Kay and Phillips first decomposed the total entropy of the output, $H(Y)$ as:
\begin{equation}
 H(Y)=I(Y:X_1:X_2)+ I(Y:X_1|X_2) + I(Y:X_2|X_1) +H(Y|X_1,X_2)~,
\end{equation}
\noindent where the three-term multi-information $I(Y:X_1:X_2)$ is defined as:
\begin{align}
\label {eq:MultiInf}
\nonumber I(Y:X_1:X_2)&= I(X_1:Y) - I(X_1:Y|X_2) \\
\nonumber &= I(X_1:X_2) - I(X_1:X_2|Y)\\
&= I(Y:X_2) - I(Y:X_2|X_1)~.
\end{align}

Kay and Phillips then re-weighted the terms of this decomposition by  coefficients $\Phi_i$ to obtain a \emph{generic} information theoretic goal function $F$ as:
\begin{equation}
\label{eq:F}
 F=\Phi_0 I(Y:X_1:X_2)+ \Phi_1 I(Y:X_1|X_2) + \Phi_2 I(Y:X_2|X_1) +\Phi_3 H(Y|X_1,X_2)
\end{equation}
 
Here, the  first term, $I(Y:X_1:X_2)$, was meant to reflect the information in the output that is shared between the two inputs, the second term the information in the output that was only in the driving input, the third term the information in the output that was only in the contextual input, while the last term represents the unused bandwidth (see Figure \ref{fig:PIDofCIP} for a graphical representation of these terms). Below, these assignments will be investigated using PID.

In previous work \cite{phillips_discovery_1995}, the goal of \emph{coherent infomax}  was implemented by setting $\Phi_0=1, \Phi_1=\Phi_2=\Phi_3=0,$ leading to the objective function $I(Y : X_1 : X_2)$. While this objective function appears not to explicitly embody any asymmetry between the influences of the $X_1$ and $X_2$ inputs, it is important to realize that the modulatory role played by the contextual input $X_2$ is expressed through the special form of activation function introduced in Phillips et al. (1995), and defined in Appendix 7.4.
The possibility of expressing this asymmetry explicitly in the objective function was also discussed in \cite{phillips_discovery_1995,kay_contextually_1998} by taking $\Phi_0 =1$, $0 \leq \Phi_1 < 1, \Phi_2 = \Phi_3=0,$ leading to the goal function
\begin{equation}
\label{eq:F_CIP}
 F_{CIP}= I(Y:X_1:X_2)+ \Phi_1 I(Y:X_1|X_2),
\end{equation}
\noindent
which is a weighted combination of the multi-information and the information between $Y$ and the driving input $X_1$ conditional on the contextual input $X_2$. This last term was meant to represent information that was both in the output $Y$ and the driving input $X_1$, but not in the contextual input $X_2$.

Next, we will investigate how this goal function $F_{CIP}$ implements the goals \ref{1}-\ref{4} when these are restated using the language of PID.

\subsection{$F$ as seen by PID}
We first take the generic goal function $F$ from equation \ref{eq:F}, that is independent of CIP proper, and rewrite it as a sum of mutual information terms and decompose these using PID. We will sort the resulting decomposition by PID terms and compare this result to the general goal function $G$. This will tell us about the space of goal functions covered by $F$. Knowing this space is highly useful as a working neural network implementation of $F$ with learning rules exists (reviewed in \cite{KayBookchapter1999,Kay2011}). This implementation can also be used to implement goal functions formulated in the precise PID framework based on $G$, whenever the specific $G$ that is of interest lies in the space that can be represented by $F$'s. 

We begin by decomposing $F$ mutual information terms:
\begin{align}
\nonumber F=&\Phi_0 I(Y:X_1:X_2) + \Phi_1 I(Y:X_1|X_2) + \Phi_2 I(Y:X_2|X_1) + \Phi_3 H(Y|X_1,X_2)\\
\nonumber =&\Phi_0 \left( I(Y:X_2) - I(Y:X_2|X_1) \right) \\
\nonumber &+ \Phi_1 \left(I(Y:X_1,X_2)-I(Y:X_2) \right)\\
\nonumber &+ \Phi_2 \left(I(Y:X_2,X_1)-I(Y:X_1) \right) \\
 &+ \Phi_3 H(Y|X_1,X_2)~,
\end{align}

\noindent which, using the PID equations \ref{eq:totalMI}-\ref{eq:condMI}, and collecting PID terms, turns into:
\begin{align}
\label{eq:PIDofF}
\nonumber F =& \Phi_1 I_{unq}(Y:X_1\setminus X_2) \\
\nonumber  &+ \Phi_2 I_{unq}(Y:X_2 \setminus X_1)\\
\nonumber  &+ \Phi_0 I_{shd}(Y:X_1;X_2)\\
\nonumber  &+ (\Phi_1+\Phi_2-\Phi_0)I_{syn}(Y:X_1;X_2) \\
  &+ (\Phi_3) H(Y|X_1,X_2)~.
\end{align}

Comparing this to the general PID goal function $G$, we see that the coefficients $\mathbf{\Gamma}=[\Gamma_0\ldots\Gamma_4]$  and $ \mathbf{\Phi}=[\Phi_0\ldots\Phi_3] $ are linked by the matrix $\Omega$ as:
\begin{align}
\label{eq:OmegaGamma}
 \Omega \mathbf{\Phi}&\eqdef \mathbf{\Gamma}  \\
 \Omega &= \left(
\begin{array}{cccc}
0 & 1 & 0 & 0\\
0 & 0 & 1 & 0 \\
1 & 0 & 0 & 0 \\
-1 & 1 & 1 & 0 \\
0 & 0 & 0 & 1 
\end{array} \right)~.
\end{align}
Since $\Omega$ is not invertible, there are parameter choices in terms of $\mathbf{\Gamma}$ that have no counterpart in $\Phi$. These are described by the complement of the range of this matrix (the null space of $\Omega^T$). This one-dimensional subspace is described by\footnote{The corresponding nullspace for the $\gamma_i$ reads: $V_{\gamma}=\{\mathbf{\gamma}\in\mathbb{R}^5:\mathbf{\gamma}=\alpha\cdot [-1,-1,1,1,0]^T,~\alpha \in \mathbb{R}\}$. }:
\begin{equation}
\label{eq:rangeComplement}
 V_{\Gamma}=\{\mathbf{\Gamma}\in\mathbb{R}^5:\mathbf{\Gamma}=\alpha\cdot [-1,-1,1,1,0]^T,~\alpha \in \mathbb{R}\}~.
\end{equation}
\noindent The existence of this subspace of coefficients not expressible in terms of $\Phi_i$'s means that it is impossible to prescribe the goal of simultaneously maximizing synergistic and shared information, while minimizing the two unique contributions, and vice versa when using $F$. Ultimately, the existence of a subspace not representable by $\Phi_i$'s is a consequence of the fact that PID terms cannot be expressed using classic information theory (while $F$ in contrast was defined from classical information theoretic terms only).

\subsection{The coherent infomax principle as seen by PID}
For the investigation of the \emph{specific} goal function $F_{CIP}$, we first want to clarify how we understand the four goals listed in the previous section. To this end we identify them one to one  with goals in terms of PID as:
\begin{enumerate}
  \item \label{CIP1byUs} $\rightarrow$ \ref{1}: The output should contain as much shared information $I_{shd}(Y:X_1,X_2)$ as possible.
  \item $\rightarrow$ \ref{2}: The output could contain some unique information $I_{unq}(Y:X_1\setminus X_2)$.
  \item  $\rightarrow$  \ref{3}: The output should minimize unique information $I_{unq}(Y:X_2\setminus X_1)$.
\item  $\rightarrow$ \ref{4}: The unused output bandwidth $H(Y|X_1,X_2)$ should be minimized.
\end{enumerate}

With respect to item \ref{CIP1byUs} on this list, it is important to recall from section \ref{sec:PID} that shared information can arise from mutual information between the sources (source shared information) or be created by a mechanism in the processor (mechanistic shared information). Kay and Phillips had in mind the first of these two possibilities.   

\begin{figure}[thpb]
\begin{center}
\includegraphics[width=\textwidth]{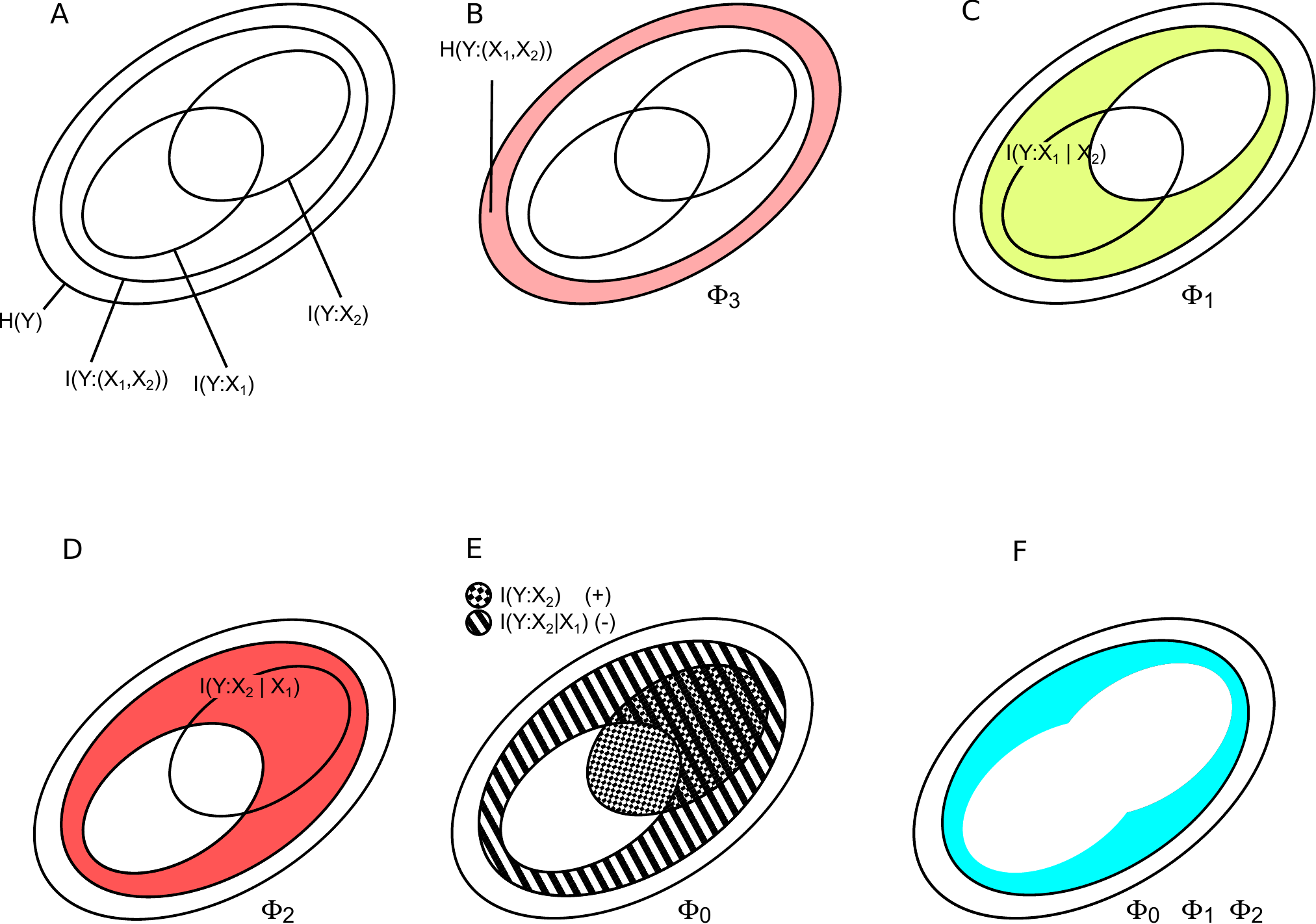}
\end{center}
\caption{ Graphical depiction of the various contributions to $F$ and their weighting coefficients in the PID diagram. (A) Classical unconditional mutual information terms. (B) Unused bandwidth, weighted by $\Phi_3$. (C) Conditional mutual information $I(Y:X_1|X_2)$, weighted by $\Phi_1$. (D) Conditional mutual information $I(Y:X_2|X_1)$, weighted by $\Phi_2$. Note the overlap of this contribution with the one from (C). (E) The three way information $I(Y:X_1:X_2)$, weighted by $\Phi_0$. Here the three way information is the checkered minus the striped area. (F) This region appears in (C),(D),(E) and is weighted accordingly by three coefficients simultaneously ($\Phi_0$,$\Phi_1$, $\Phi_2$). The area in (F) is the synergistic mutual information that is also shown in cyan in Fig. \ref{fig:PID}.}
\label{fig:PIDofCIP} 
\end{figure}

To see whether $F_{CIP}$ indeed reflects these goals as stated via PID, we look at the specific choice of parameters, $\Phi_0=1$, $ 0 \leq \Phi_1 < 1$, $\Phi_2= \Phi_3 =0$, that was used to implement the coherent infomax principle, and find using equations \ref{eq:totalMI}-\ref{eq:condMI} (the reader may also verify this graphically using Figure \ref{fig:PIDofCIP}):
\begin{align}
\label{eq:PIDofCIP_2}
F_{CIP} = 
   I_{shd}(Y:X_1;X_2) + \Phi_1 I_{unq}(Y:X_1\setminus X_2)  - (1 -\Phi_1) I_{syn}(Y: X_1;X_2).
\end{align}

We will now discuss the various contributions to $F_{CIP}$ in detail, starting with the shared information, which figures most prominently in the goals \ref{1}-\ref{4}.

\paragraph{Shared information} We see that shared information is  maximized. This shared information contains contributions from mutual information between the sources (source shared information) as well as shared information created by mechanisms in the processor (mechanistic shared information, see the note on item \ref{CIP1byUs} above). The first type of shared information is the one aimed for in \ref{1}. Thus, for inputs that are not independent the coherent infomax goal function indeed maximizes source shared information as desired. We will investigate the case of independent inputs below.

\paragraph{Unique information} In addition to the shared information, the unique information from the driving input is also maximized, albeit to a lesser degree. In contrast, synergy between the output and the combined inputs is minimized. Therefore, goals 1, 2 and 3 are expressed explicitly in this objective function but there is no explicit mention of minimizing the output bandwidth.

\paragraph{Synergistic information} Of all the PID terms, synergy is discouraged.  This may at first seem surprising as the mapping of goals of coherent infomax to PID, did not appear to make any explicit statements about synergistic components -- \textit{unless} one views the transmission of undesirable synergistic components as being an extra component of the bandwidth (along with $H(Y|X_1,X_2)$) that is not used in the optimal attainment of goals 1-3. Nevertheless the minimization of synergy serves the original goals of coherent infomax. This can be seen when we consider that these were formulated for two different types of inputs, driving and modulatory. For these two types of input, the goal of coherent infomax is to use the modulatory inputs to guide transmission of information about the driving inputs. Synergistic components would transmit information about both driving and modulatory inputs, so transmitting them would be treating the modulatory inputs as driving inputs. This is clearly undesirable in the setting of coherent infomax.

At a more technical level, we note the trade-off in that increasing the value of the parameter $\Phi_1$ towards 1 at once serves to enhance promotion of the unique information from the driving input while simultaneously lessens the pressure to minimize the synergy. This is a remnant of the term $\Phi_1 I(Y:X_1|X_2)$ in \eq{F_CIP} which had been included in order to capture information that was both in $Y$ and $X_1$ but not in $X_2$ (i.e. the unique information from the driving input), but inadvertently also served to capture the synergy.

In terms of the range of tasks that can be learned by a processor with $F_{CIP}$, the minimization of synergy between the two types of inputs means for example that learning tasks that require a lot of synergy between the inputs, like the \texttt{XOR}-function, cannot be achieved easily. It is crucial, however, to realize that discouragement of synergy concerns only relations between drive $X_1$ and modulation $X_2$. In contrast, synergistic relations between just the components of a multivariate $\mathbf{X_1}$ can be learned by the coherent infomax learning rule. The \texttt{XOR} between components of $\mathbf{X_1}$ for example can be learned reliably if supervised, and still occasionally if not \cite{phillips_discovery_1995}.

 \paragraph{Independent sources}  What remains to be investigated is what the goal functions aims for in the specific case of statistically independent inputs, i.e. when source shared information cannot be obtained. In other words, we may ask whether the coherent infomax processor will maximize mechanistic shared information in this case?

Since the mutual information between the inputs, $I(X_1:X_2)$, is assumed to be zero, then using one of the forms of the multi-information (eq. \ref{eq:MultiInf}) we have
\begin{equation}
 I(Y:X_1:X_2) = I(X_1:X_2) - I(X_1:X_2|Y) = - I(X_1:X_2|Y)
\end{equation}
and so the multi-information is non-positive.  It follows from the other forms of the multi-information (eq. \ref{eq:MultiInf}) that
\begin{equation}
  I(Y:X_1|X_2) \geq I(Y:X_1) \quad \text{and} \quad I(Y:X_2|X_1) \geq I(Y:X_2).
\end{equation}

This implies directly (compare \ref{fig:PID}A) that for independent inputs we must have:
\begin{equation}
I_{shd}(Y:X_1; X_2) \leq I_{syn}(Y:X_1; X_2)~,
\end{equation} 
\noindent -- an important additional constraint that arises from independent inputs. Thus, in this case the minimization of synergy and the maximization of shared information compete, giving more effective weight to the unique information from the driving input. Nevertheless, limited shared information may exist in this scenario, and if so it will be of the mechanistic type.

In sum, we showed that (i) the generic goal function $F$ in the coherent infomax principle cannot represent all goal functions that are possible in the PID framework using the goal function $G$ -- specifically, $F$ lacks one degree of freedom; (ii) for the CIP this leads to a weighted maximization of the shared information (source shared information and mechanistic shared information) and the unique information from the driving input; (iii) it can be shown that within the space of all possible goal functions $F$ it is impossible to maximize synergy and shared information together, while minimizing the two unique information terms, and vice versa; (iv) and for the CIP synergy between the driving and modulatory inputs is explicitly discouraged.

\section{Partial information decomposition as a unified framework to generate neural goal functions}
In the this section we will use PID to investigate infomax, another goal function proposed for neural systems, and we will formulate an information-theoretic goal function for a neural processor aimed at predictive coding.

\subsection{Infomax}
To investigate infomax, we recall that the goal stated there is to maximize the information in the output about the relevant input $X_1$, which typically is multivariate \cite{Linsker1988}. This goal function is implicitly designed for situations with limited output bandwidth, i.e. $H(X_1)>H(Y)$. Not considering a second type of input $X_2$ it is obvious that PID will not contribute to the understanding of infomax. This changes however if the variables in a multivariate input will be considered  separately. Then, it may make sense to ask whether the output information in a given system is actually being maximized predominantly due to unique or synergistic information.

Mathematically, the infomax goal can also be represented by using $F$ with two types of inputs $X_1$, $X_2$, where the information transmitted about $X_1$ is to be maximized. This can be achieved by choosing $\Phi_0=\Phi_1=1$ to obtain (e.g. \cite{KayBookchapter1999}):

\begin{align}
 \label{eq:Infomax}
 F_{IM}=&\Phi_0 I(Y:X_1:X_2) + \Phi_1 I(Y:X_1|X_2) \\
 \nonumber =& I_{unq}(Y:X_1 \setminus X_2)+I_{shd}(Y:X_1;X_2)\\
 \nonumber =& I(Y:X_1)~.
\end{align}
\noindent The insight to be gained using PID here is that infomax does not incorporate the use of auxiliary variables $X_2$ to extract even more information from $X_1$ via the synergy $I(Y:X_1;X_2)$, nor does it prefer either shared or unique information over the other.

\subsection{Predictive coding}
In predictive coding the goal is to predict inputs $X_1(t)$ using information available from past inputs $\mathbf{X_1(t-1)}=[X_1(t-1) \ldots X_1(t-k)]$\footnote{These past inputs may in principle lie arbitrarily far in the past (i.e. with arbitrarily large $k$), meaning that also long term memory in a system may contribute to the predictions.}. Thus, the processor has to learn a model $M_{PC}$ that yields predictions $X_2(t)=M_{PC}(\mathbf{X_1(t-1)})$, such that $X_2(t)\approx X_1(t)$. This is the same as maximizing the mutual information between outcome and prediction $I(X_1(t),X_2(t))=I(X_1(t),M_{PC}(\mathbf{X_1(t-1)}))$, at least if we do not care how exactly $X_2(t)$ \emph{represents}\footnote{Here, representation is used in the sense of lossless encoding. Thus, for us $X_2(t)$ is equivalent to all lossless (re-)encodings of $X_2(t)$, e.g. in other alphabets, amongst others the alphabet of $X_1(t)$.} the prediction. Under some mild constraints \footnote{The relevant constraints here are that the collections of input values $\mathbf{X_1(t)}$ are sampled appropriately, such that they form a Markov chain $\mathbf{X_1(t-2)}\rightarrow \mathbf{X_1(t-1)} \rightarrow \mathbf{X_1(t)}$} the data processing inequality here actually states that trying to tackle this problem information theoretically is trivial, as $I(X_1(t),X_2(t))=I(X_1(t),M_{PC}(\mathbf{X_1(t-1)}))$ is maximized by 
$M_{PC}(\mathbf{X_1(t-1)})=\mathbf{X_1(t-1)}$, i.e. all the information we can ever hope to exploit for prediction is already in the raw data (and it is a mere technicality to extract it in a useful way). The whole problem becomes interesting only when there is some kind of bandwidth limitation on $M_{PC}$, i.e. when  for example $M_{PC}(\mathbf{X_1(t-1)})$ has to use the same alphabet as $X_1(t)$, meaning that we have to state our prediction as a single value that $X_1(t)$ will take. Of course, this actually is the typical scenario in neural circuits. Therefore, we state the main goal of predictive coding as maximizing $I(X_1(t),X_2(t))=I(X_1(t),M_{PC}(\mathbf{X_1(t-1)}))$, under the constraint that $X_1(t)$ and $M_{PC}(\mathbf{X_1(t-1)}))$ have the same ``bandwidth'' (the same raw bit content to be precise). Despite of the goal to maximize a simple mutual information this is not an infomax problem, due to the temporal order of the variables, i.e. we need the output $X_2(t)$ before the input $X_1(t)$ is available. Thus, we have to find a different solution to our problem.

\begin{figure}[thpb]
\begin{center}
\includegraphics[width=\textwidth]{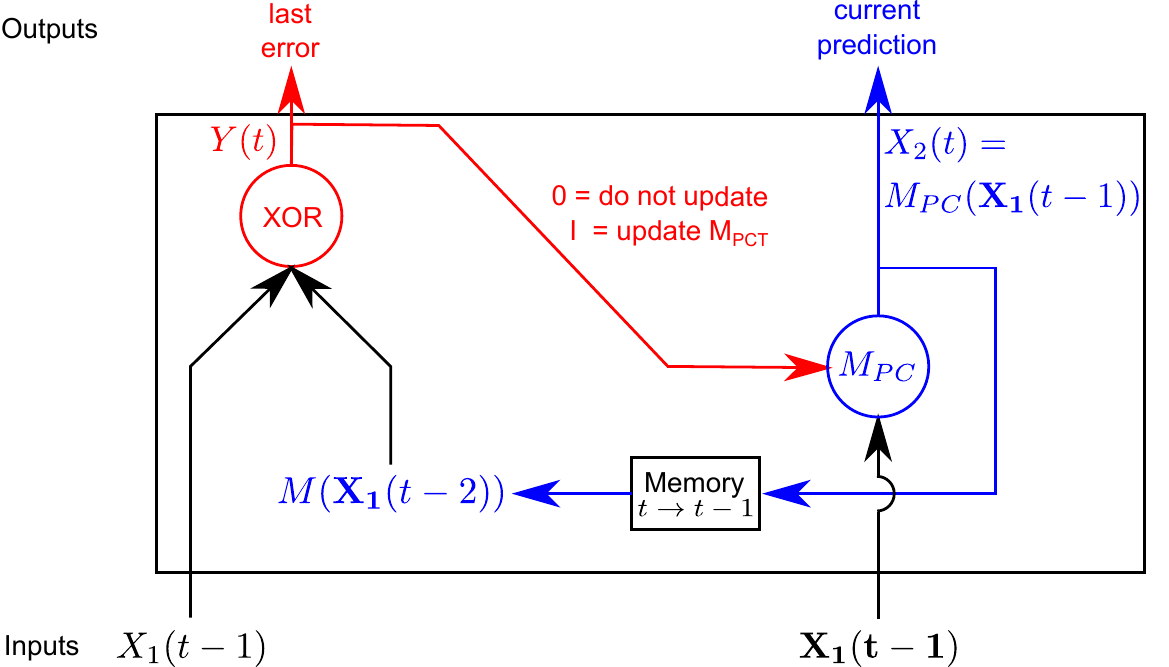}
\end{center}
\caption{ Graphical depiction of a minimalistic, binary predictive coding circuit. This circuit can be conceived of as \emph{one} neural processor (indicated by the box) with inputs $X_1(t-1)$, $\mathbf{X_1}(t-1)$ and (main) output $Y(t)$.}
\label{fig:PredictiveCodingCircuit_minimal} 
\end{figure}

To this end, we suggest that a minimal circuit performing predictive coding will have to perform at least three subtasks, (i) produce predictions as output, (ii) detect whether there were errors in the predictions, (iii) use these for learning. In Fig.\,\ref{fig:PredictiveCodingCircuit_minimal} we detail a minimalistic circuit performing these tasks, with subtask (i) represented in $X_2(t)$, subtask (ii) in $Y(t)$ and subtask (iii) in $M_{PC}$. This circuit assumes the following properties for its neural circuits: (a) neurons have binary inputs and outputs, (b) information passes through a neuron in one direction, and (c) information from multiple inputs can be combined into one output only. The circuit consists of two separate units: (1) the error detection unit that operates on \emph{past} predictions $X_2(t-1)=M_{PC}(\mathbf{X_1(t-2)})$, obtained via a memory buffer, and past inputs $X_1(t-1)$, to create the output $Y$ via an \texttt{XOR} operation, with $y=1$ indicating an erroneous prediction in the past; (2) the prediction unit that has the capability to produce output based on a weighted summation over a vector of past inputs $\mathbf{X_1(t-1)}$ via a weighting function in the model $M_{PC}$.  $M_{PC}$ will update its weights whenever an error was received.

We suggest that the information theoretic goal function of this circuit is simply to  minimize the entropy of the output of the error unit, i.e. $H(Y)$. In principle, this would drive the binary output of the circuit either to $p(y=1)\rightarrow 1$ or to $p(y=0)\rightarrow 1$. Of these two possibilities, only the second one is stable, as the constant signaling of the presence of an error will lead to incessant changes in $M_{PC}$, which in turn will change $Y$ even for unchanging input $X_1$. Thus, minimizing $H(Y)$ should enforce $p_Y(y=0)\rightarrow 1$. Therefore, we can formulate an information theoretic goal function of the form $G$ if we conceive of the whole circuit as being just \emph{one} neural processor with inputs $X_1(t-1)$ and $\mathbf{X_1(t-1)}$, and as having the error $Y$ as its main output. In this case, we find as a goal function for the predictive coding error (PCE):
\begin{align}
G_{PCE} =& \gamma_0 I_{unq}(Y:X_1(t-1) \setminus \mathbf{X_1}(t-1)) \\
\nonumber &+ \gamma_1 I_{unq}(Y:\mathbf{X_1}(t-1) \setminus X_1(t-1)) \\
\nonumber &+ \gamma_2 I_{shd}(Y: X_1(t-1);\mathbf{X_1}(t-1)) \\
\nonumber &+ \gamma_3 I_{syn}(Y: X_1(t-1);\mathbf{X_1}(t-1)) \\
\nonumber &+ \gamma_4 H(Y) ~,
\end{align}
\noindent with the weights $\mathbf{\gamma}_{PCE}=[0,0,0,0,-1]$ using the $\gamma$-notation from equation \ref{eq:GoalFunction_HY} where the total output entropy was made explicit, or equivalently, $\mathbf{\Gamma}_{PCE}=[-1,-1,-1,-1,-1]$.
Interestingly, this goal function formally translates to $\mathbf{\Phi}=[-1,-1,-1,-1]$, or $F=-H(Y)$. This gives hope that one can translate the established formalism for $F$ to the present case by taking into account that the original architecture behind $F$ is augmented here by an additional \texttt{XOR} subunit.
Learning of the circuit's goal function may have to proceed in two steps if we do not have subunits able to perform \texttt{XOR} at the beginning. In this case, the ``\texttt{XOR}'' subunit will first have to learn to perform its function. This can be achieved by maximizing the synergy of two uniform, random binary inputs and the subunit's output $Y$. After this initial learning the \texttt{XOR}-subunit is `frozen' and learning of predictions can proceed to minimize $H(Y)$. One conceivable mechanism for this would be to use learning based on coincidences between input bits in $M(\mathbf{X_1}(t-2))$ and the error bit $Y$.

We note that this goal function is not entirely new,  as the idea of making the output of a processing unit as constant as possible in learning  has been used before in various implementations (e.g. \cite{Wyss2006,Cannon1932,Der1999}). It is also closely related to the homeostatic goals pursued by the free energy minimization principle \cite{Friston2006a,Friston2007, Friston2009b}.
We have merely added here a generic minimal circuit diagram and the information theoretic interpretation to these previous approaches. Also, note that the actual prediction $X_2(t)=M_{PC}(\mathbf{X_1(t-1)})$ must be implicitly part of the information theoretic goal function, as the goal function we suggest here would be nonsensical on many other circuits.

As a next level of complication one may consider that the predictions $X_2$ that are created within our minimal circuit are sent back to the source of the input $X_1$ to interact with it there. One such interaction scheme will be studied in the next section.

\section{Coding with synergy}
So far the goal functions investigated in our unifying framework $G$ had in common that maximization of synergy did not appear as a desirable goal. This may historically be simply due to the profound mathematical difficulties that had to be overcome in the definition of synergistic information. In this section we will therefore show how synergy naturally arises in a generalization of ideas from efficient coding by PID. We will call the goal function simply \emph{coding with synergy (CWS)}.

The neural coding problem that we will investigate here is closely related to predictive coding discussed in the previous section. However, in contrast to predictive coding where the creation of predictions was in focus, here we focus on possible uses of prior (or contextual) information from $X_2$, be it derived from predictions or by any other means. In other words, we here simply assume that there is (valid) prior information in the system that does not have to be extracted from the ongoing input stream $X_1$ by our neural processor. Moreover, we assume that there is no need to waste bandwidth and energy on communicating $X_2$ as this information is already present in the system. Last, we assume that we want to pass as much of the information in $X_1$ as possible, as well as of the information created synergistically by $X_1$ and $X_2$. This synergistic information will arise for example when $X_2$ serves to decode or disambiguate information in $X_1$.
 
 Looking at the PID diagram (Fig. \ref{fig:PID}) one sees that in this setting it is optimal to minimize $I_{unq}(Y:X_2 \setminus X_1)$ and the unused bandwidth $H(Y|X_1,X_2)$ while maximizing the other terms. This leads to:
 
 \begin{align}
 \label{eq:CWS}
  G_{CWS} =& I_{unq}(Y:X_1\setminus X_2) \\
\nonumber &- I_{unq}(Y:X_2 \setminus X_1) \\
\nonumber &+ I_{shd}(Y: X_1;X_2) \\
\nonumber &+ I_{syn}(Y: X_1;X_2) \\
\nonumber &- H(Y|X_1,X_2) ~,
 \end{align}
 \noindent or $\mathbf{\Gamma}=[1, -1, 1, 1, -1]$. The important point here is that this is different from maximizing just $I(Y:X_1 | X_2)$, as this would omit the shared information, 
i.e. we would lose this part of the information in $X_1$. The goal function $G_{CWS}$ is also different from just maximizing $I(Y:X_1)$, as this would omit the synergistic information, i.e. the possibility to decode information from $X_1$ by means of $X_2$.
Furthermore, there is no corresponding goal function $F$ here in terms of classical information theoretic measures. This can easily be proven by noting that $\mathbf{\Gamma}=[1, -1, 1, 1, -1]$ has a non-zero projection in $V_\Gamma$ (\eq{rangeComplement}). In other words, there is no $\mathbf{\Phi}$ that satisfies equation \ref{eq:OmegaGamma}.
 
  Given there were bandwidth constraints on $Y$, one might want to preferentially communicate one or two of the positively weighted terms in equation \ref{eq:CWS}. The natural choice here is to favor synergy and unique information about $X_1$, because the shared information with $X_2$ is already in the system. If just one contribution can be communicated this leaves us with three choices. We will quickly discuss the meaning of each here: first, focusing on the unique information $I_{unq}(Y:X_1 \setminus X_2)$ emphasizes the surprising information in $X_1$, because this is the information that is not yet in the system at all (i.e. not in $X_2$); second, focusing on the shared information $I_{shd}(Y: X_1;X_2)$ basically leads to coherent infomax; third, focusing on the synergistic information  $I_{syn}(Y: X_1;X_2)$ emphasizes information which can only be obtained when putting together prior knowledge in $X_2$ and incoming information $X_1$ - this would be the extreme case of CWS. This case should arise naturally in binary error computation, e.g. in error units suggested as integral parts of certain predcitive coding architectures (see \cite{clark_whatever_2013} for a discussion of error units, also compare the XOR unit in Figure \ref{fig:PredictiveCodingCircuit_minimal}).
 
 A classic example for this last coding strategy would be cryptographic decoding. Here, the mutual information between cypher text (serving as input $X_1$) and plain text (serving as output $Y$) is close to zero, i.e. $I(Y:X_1)\approx 0$, given randomly chosen keys and a well performing cryptographic algorithm. Nevertheless the mutual information between the two, given keys (serving as input $X_2$), is the full information of the plain text, i.e. $I(Y:X_1|X_2)=H(Y)$, assuming the unused bandwidth is zero ($H(Y:X_1,X_2)=0$). As the mutual information between key and plain text should also be zero ($I(Y:X_2)=0$)  we see that in this case the full mutual information is synergistic: $I(Y:X_1,X_2)=I_{syn}(Y:X_1;X_2)$. 
 In a similar vein, any task  in neural systems that involves an arbitrary key-dependent mapping between information sources -- as in the above cryptographic example -- will involve CWS. One such task would be to read a newspaper printed in Latin characters (which could be in quite a range of languages) to get knowledge about the current state of the world (or at least some aspects of it). Visually inspecting the text, without the information incorporated in the rules of the unknown written language used will not reveal information about the world. Yet, having all the information on the rules of written language, without having a specific text will also not reveal anything about the world. To obtain this knowledge we need, both, the text of the newspaper and the language-specific information how written words map to possible states of the world.
 
 A corollary of the properties of synergistic mutual information is that when a neuron's inputs are investigated  individually they will seem unrelated to the output -- to the extent that synergistic information is transmitted in the output. Therefore, the minimal configuration of neuronal recordings needed to investigate the synergistic goal fucntion is a triplet of two inputs and one output. Thus, though coding with synergy has not been prominent in empirical reports to date, it might become more frequently detected as dense and highly parallel recordings of neuronal acticity become more widely available.

 The general setting of coding under prior knowledge discussed here is also related to Barlow's efficient coding hypothesis \cite{Barlow1961} if we take the prior information $X_2$ to be information about which inputs to our processor are typical for the environment it lives in. We here basically generalize Barlow's principle by dropping reference to what the input or the prior knowledge are about. 
 
 Last, this goal function seems significant to us as synergy is seen by some authors as useful in an formal definition of information modification (e.g. \cite{Lizier2013InfoMod}). Thus synergy is a highly useful measure in the description of neural processor with two or more inputs (or one input and an internal state), as it taps into the potential of the processor to genuinely \emph{modify} information \footnote{Most interestingly, the current definition of information transfer via the transfer entropy \cite{Schreiber2000} actually also contains an element of synergy between the source's and the target's past \cite{Williams2011TE}, and thus there are basically modifying and non-modifying forms of information transfer.}
 
\section{Discussion}
\subsection{Biological neural processors and PID}
In this study we introduced partial information decomposition (PID) as a universal framework to describe and compare neural processors in a domain-independent way. PID is indispensable for the information theoretic analysis of systems where two (or more) inputs are combined to one output, because it allows to decompose the information in the output into contributions provided either uniquely by any one of the inputs alone (unique information), by either of them (shared information), or only by both of them jointly (synergistic information). Using PID, the information processing principles of the processor can be quantitatively described by specific coefficients $\mathbf{\Gamma}$ for each of the PID contributions in a PID-based goal function $G(\mathbf{\Gamma})$, which the processor maximizes. 

This framework is useful in several ways. First, and perhaps most importantly, it allows the principled comparison of existing neural goal functions, such as infomax, coherent infomax, predictive coding, and efficient coding. Second, it aids in the design of novel neural goal functions. Here we presented a specific example, coding with synergy (CWS), that exploits synergy to maximize the information that can be obtained from the input when prior information is available in the system. Note, however, that the actual implementation of a neural circuit maximizing the desired goal function is not provided by the new framework and will have to be constructed on a case by case basis at the moment. This is in contrast to coherent infomax where a working implementation is known. Third, applying this framework to neural recordings may help us understand better how neural circuits that are far away from sensory and motor periphery, and for which we do not have the necessary semantics, function.

Currently, the applicability of our framework rests on the assumption that a neural processor with \textit{two} inputs is a reasonable approximation of a neuron or microcircuit \footnote{Although the work of Griffiths \cite{Griffith2014a} and colleagues as well as Bertschinger and colleagues \cite{Rauh2014} allows some extensions to more inputs and outputs, respectively.}. Of course, neurons typically have many more inputs than just two. However, if such inputs naturally fall into two groups, e.g. being first integrated locally in two groups on the dendrites before being brought to interact at the soma, then indeed the two input processor is a useful approximation. If, moreover, these integrated inputs are measured before their fusion in the soma, then the formalism of goal functions presented here will allow us to assess the function of this neuron in a truly domain independent way, relying only on information that is also available to the neuron itself.

For example, two such spatially segregated and separately integrated inputs can be distinguished on Pyramidal cells (Fig. \ref{fig:NeuralProcessor}). Pyramidal cells are usually highly asymmetric and consist of a cell body with basal dendrites and an elongated apical dendrite that rises to form a distal dendritic tuft in the superficial cortical layers. Thus, the inputs are spatially segregated into basal/perisomatic inputs, and inputs that target the apical tuft. Intracellular recordings indicate that there are indeed separate integration sites for each of these two classes of input, and that there are conditions in which apical inputs amplify (i.e. modulate) responses to the basal inputs in a way that closely resembles the schematic two-input processor shown in Fig. \ref{fig:NeuralProcessor}. There is also emerging evidence that these segregated inputs have driving and modulatory functions and are combined in a mechanism of apical amplification of basal inputs -- resembling the coherent infomax goal function. Direct and indirect evidence on this apical amplification and its cognitive functions is reviewed by Phillips [submitted to this special issue]. That evidence shows that apical amplification occurs within pyramidal cells in the superficial layers, as well as in layer 5 cells, and suggests that it may play a leading role in the use of predictive inferences to modulate processing.

Which of the goal functions proposed here, e.g infomax, coherent infomax, or coding with synergy a neural processor actually performs is an empirical question that must be answered by analyzing PID footprints of $G$ obtained from data recorded in neural processors. At present this is still a considerable challenge when applied to the level of single cells or microcircuits because this requires the separate recording of at least one output and two inputs, wich must moreover be of different type in the case of coherent infomax. Next, the PID terms have to be estimated from data, instead of distributions that are known. This type of estimation is still a field of ongoing research at present. Overcoming these challenges will yield in-depth understanding of, for example, the information processing of the layer 5 cell described above in terms of PID, and elucidate which of the potential goal functions is implemented in such a neuron.

In the spirit of the framework proposed here, classical information theoretic techniques have already been applied to psychophysical data to search for coherent infomax-like processing at this level \cite{phillips_interactions_2000}. These studies confirmed for example that attentional influences are modulatory, and showed how modulatory interactions can be distinguished from interactions that integrate multiple driving input streams. These result are a promising beginning of a more large scale analysis of neuronal data at all levels with information theoretic tools, such as PID.

Further information theoretic insight relevant to predictive processing may also be gained by relating the predictable information in a neural processor's inputs (measured via 'local active information storage' \cite{Lizier2012a}) to the information transmitted to its output (measured via transfer entropy \cite{Schreiber2000}, or local transfer entropy \cite{Lizier2008}) to investigate whether principles of predictive coding apply to the information processing in neurons. This is discussed in more detail in \cite{wibral_bits_2015}.

\subsection{Conclusion}
We here argued that the understanding of neural information processing will profit from taking a neural perspective, focusing on the information entering and exiting a neuron, and stripping away semantics imposed by the experimenter -- semantics that is not available to a neuron. We suggest that the necessary analyses are best carried out in an information theoretic framework, and that this framework must be able to describe the processing in a multiple input system to accommodate neural information processing. We find that PID provides the necessary measures, and allows to compare most if not all theoretically conceivable neural goal functions in a common framework. Moreover, PID can also be used to design new goal functions from first principles. We demonstrated the use of this technique in understanding neural goal functions proposed for the integration of contextual information (coherent infomax), the learning of predictions (predictive coding), and introduced a novel one for the decoding of input based on prior knowledge called coding with synergy (CWS).

\section*{Acknowledgements}
The authors would like to thank Nils Bertschinger for inspiring discussions on partial information decomposition and for reading an earlier version of the manuscript. MW received support from LOEWE Grant "Neuronale Koordination Forschungsschwerpunkt Frankfurt (NeFF)". VP  received financial support from the German Ministry for Education and Research (BMBF) via the Bernstein Center for Computational Neuroscience (BCCN) G\"{o}ttingen under Grant No. 01GQ1005B.


\begin{appendix}
\section{Notation}
\label{app:Notation}
\subsection{Probability distributions}
We write probability distributions of random variables $X_1$, $X_2$, $Y$ as $P(X_1,X_2,Y)$ wherever we're talking about the distribution as an object itself, i.e. when we treat a distribution $P(X_1,X_2,Y)$ as a point of in the space of all joint probability distributions of these three variables. To signify a value that such a distribution takes for specific realizations $x_1$, $x_2$, $y$ of these variables, we either write $P(X_1=x_1, X_2=x_2, Y=y)$, or use the shorthand $p(x_1,x_2,y)$. 
\subsection{Notation of PID terms}
To highlight the necessity of the notation used here and to deepen the understanding of the various partial information terms we give the following example where we add explicit set notation for clarity:
\begin{align}
 \label{eq:PIDallshared}
 I_{shd}(Y:A;B;C;D)\neq\\
 \label{eq:PIDoneset}
 I_{shd}(Y:A,B,C,D) = \\ 
 \nonumber  I_{shd}(Y:\{A,B,C,D\})=\\ 
 \nonumber I_{shd}(Y:\{A,B,C,D\};\{A,B,C,D\})=\\
 \nonumber I(Y:A,B,C,D),
\end{align}
\begin{align}
\label{eq:PIDfoursets}  
I_{shd}(Y:A;B;C;D)\neq\\
 \label{eq:PIDtwosets}  I_{shd}(Y:A,B;C,D) = \\
 \nonumber  I_{shd}(Y:\{A,B\};\{C,D\})~.
\end{align}
\noindent Here, the first expression (\ref{eq:PIDallshared}) asks for the information that \emph{all} four right hand side variables share about $Y$, while the second expression (\ref{eq:PIDoneset})  asks for the information that the set $\{A,B,C,D\}$ shares (with itself) about $Y$. By the self-redundancy axiom \footnote{The self-redundancy axiom states that the shared information is just the mutual information between input and output when considering the same input twice, i.e. $I_{shd}(Y:X_1;X_1)=I(Y:X_1)$. This axiom becomes important in extensions of PID to more than two input variables.} \cite{PaulWilliamsPID} this is just the mutual information between the set $\{A,B,C,D\}$ and $Y$. In the next example in equations \ref{eq:PIDfoursets} and \ref{eq:PIDtwosets} we ask in equation \ref{eq:PIDtwosets} for the information shared between the two sets of variables $\{A,B\}$ and $\{C,D\}$, meaning that the information about $Y$ can be obtained from either $A$, or $B$, or from them considered jointly, but must also be found in either $C$ or $D$ or in the two of them considered jointly. This means in the latter case information held jointly by $A$ and $B$ about $Y$ is considered if it is shared with information about $Y$ obtained from any combination of $C$, $D$, including their synergistic information.

\section{Partial Information Decomposition}
\label{app:PID}
\subsection{Partial information decomposition based on unique information}
We here present a definition of unique information given by Bertschinger et al. \cite{Bertschinger2014}, which is equivalent to that provided by Griffith and Koch \cite{Griffith2014a}. We assume (that neural signals can be described by) discrete random variables $X_1$, $X_2$, $Y$ with (finite) alphabets $\mathcal{A}_{X_1}=\{{x_1}_1,\ldots,{x_1}_M \}$, $\mathcal{A}_{X_2}=\{{x_2}_1,\ldots,{x_2}_N\}$, $\mathcal{A}_Y=\{y_1,\ldots,y_L\}$, described by their joint probability distribution $P(X_1,X_2,Y)=\{p({x_1}_1,{x_2}_1,y_1),\ldots,p({x_1}_M,{x_2}_N,y_L)\}$.

As already mentioned above, a definition of either unique, or shared, or synergistic information that $X_1$, $X_2$ and $\{X_1,X_2\}$ have about a variable $Y$ is enough to have a well defined PID. Among these possibilities, Bertschinger and colleagues opt for a definition of unique information based on the everyday notion that having unique information about $Y$ implies that we can exploit this unique information to our favor against others who do not have this information -- at least given a suitable situation. Thus if we are allowed to construct such a suitable situation to our liking, we may prove to have unique information for example by winning bets on the outcomes of $Y$, where the bets are constructed by us against an opponent who does not have that unique information.

More formally, one can imagine two players $A$lice and $B$ob. Alice has access to the variable $X_1$ from equation \ref{eq:totalMI}, while she does neither have access to variable $X_2$, nor direct access to variable $Y$.  Bob has access to the variable $X_2$, but neither direct access to $X_1$, nor to $Y$. To the extent that the mutual information terms $I(Y:X_1)$, and $I(Y:X_2)$ allow, Alice and Bob however, do have \emph{some} information about the variable $Y$, despite not having direct access to $Y$.
If Alice wants to prove that having access to $X_1$ gives her unique information \emph{about} $Y$\footnote{Remember that the unique information is part of a decomposition of a mutual information, $I(Y:X_1,X_2)$, so we're looking at information \emph{about $Y$}. We do not care how much information $X_1$ has about $X_2$, and vice versa.}, then she can suggest to Bob to play a specific game, designed by her, where the payout depends only on the outcomes of $Y$. In such a game, her reward will depend only on the probability distribution $p(x_1,y)=p(x_1|y)p(y)$, while Bob's reward will depend only on $p(x_2,y)=p(x_2|y)p(y)$. The winner is thus determined simply by the two distributions $p(x_1,y)$ and $p(x_2,y)$, but not by the details of the full distribution $p(x_1,x_2,y)$. Practically speaking, Alice should therefore construct the game in such a way that her payout is high for outcomes $y$ about which she can be relatively certain, knowing $x_1$.

\begin{figure}[thpb]
\begin{center}
\includegraphics[height=0.9\textheight]{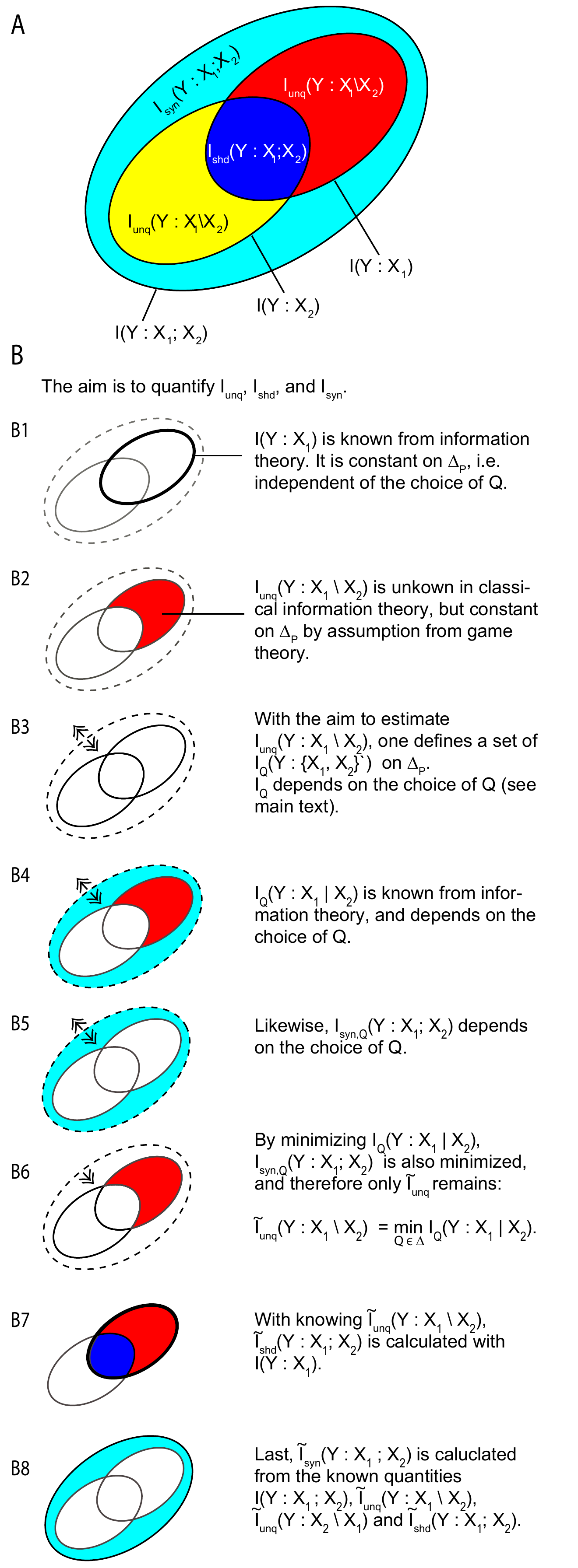}
\end{center}
\caption{Graphical depiction of the principle behind the definition of unique information in \cite{Bertschinger2014}. For details also see the main text. (A) Reminder of the partial information diagram. (B) Explanation how unique information can be defined using minimization of conditional mutual information on the space of probability distributions $\Delta_P$ (see text). Note that if the synergy in (B6) can not be reduced to 0, then we simply \emph{define} the unique information measure as $\TUNQ(Y:X_1 \setminus X_2) =\underset{Q \in \Delta_P}{min}I(Y:X_1|X_2)$.}
\label{fig:PID_full} 
\end{figure}

From this argument, it follows that Alice could not only prove to have unique information in the case described by the full joint distribution $P=P(X_1,X_2,Y)$, but also for all other cases described by distributions $Q=Q(X_1,X_2,Y)$ that have the same pairwise marginal distributions, i.e. $p(x_1,y)=q(x_1,y) \wedge p(x_2,y)=q(x_2,y) ~ \forall x_1, x_2 ,y \in \mathcal{A}_{X_1,X_2,Y}~$. Based on this observation it makes sense to request that $I_{unq}(Y:X_1 \setminus X_2)$ and $I_{unq}(Y:X_2\setminus X_1)$ stay constant on a set $\Delta_P$ of probability distributions that is defined by:
\begin{equation}
\begin{split}
 \Delta_P=\{Q \in \Delta :Q(X_1=x_1, Y=y)=P(X_1=x_1,Y=y)\\
 \text{and} Q(X_2=x_2, Y=y)=P(X_2=x_2,Y=y) \}~
 \end{split}
\end{equation}
\noindent where $\Delta$ is the set of all joint probability distributions of $X_1$, $Y$, $X_2$.

From this, it follows from equation \ref{eq:singleMI} that also the shared information $I_{shd}(Y:X_1;X_2)$ must be constant on $\Delta_P$ (consult Figure \ref{fig:PID_full}, and take into account that the mutual information terms $I(Y:X_1)$ and $I(Y:X_2)$ are also constant on $\Delta_P$). Hence, the only thing that may vary when exchanging the distribution P, for which we want to determine the unique information terms,for another distribution $Q\in\Delta_P$ is the synergistic information $I_{syn}(Y:X_1;X_2)$. It therefore makes sense to look for a specific distribution $Q_0\in\Delta_P$ where the unique information terms coincide with something computable from classic information theory. From Figure \ref{fig:PID} we see that for the case of a distribution $Q_0\in\Delta_P$ where synergistic information vanishes, the unique information terms would coincide with conditional mutual information terms, i.e. $I_{{unq},P}(Y:X_1 \setminus X_2)=I_{{unq},{Q_0}}(Y:X_1 \setminus X_2)=I_{Q_0}(Y:X_1|X_2)$. It is known, however, that a $Q_0$ with this property does not necessarily exist for all definitions of unique, shared and synergistic information that satisfy equations \ref{eq:totalMI}-\ref{eq:condMI}, and that also satisfy the above game-theoretic property (being able to prove the possession of unique information). Therefore, Bertschinger and colleagues suggested to define a measure $\TUNQ$ of unique information via the following minimization:
\begin{align}
\label{eq:TI_{unq}}
\TUNQ(Y:X_1\setminus X_2) = \min_{Q\in\Delta_{P}} I_{Q}(Y:X_1|X_2)~{}\\
\TUNQ(Y:X_2\setminus X_1) = \min_{Q\in\Delta_{P}} I_{Q}(Y:X_2|X_1)~.
\end{align}

From this, measures for shared and synergistic information can be immediately obtained via equations \ref{eq:singleMI}, \ref{eq:totalMI} as:
\begin{align}
\label{eq:TI_{shd}}
\begin{split}
 \TSHD(Y:X_1;X_2)  & =\max_{Q\in\Delta_{P}} \left( I(Y:X_1)-I(Y:X_1|X_2)\right) \\
 &= \max_{Q\in\Delta_{P}} CoI_{Q}(Y;X_1;X_2)~,
 \end{split}\\
 \label{eq:TI_{syn}}
 \TSYN(Y:X_1;X_2) &= I(Y:(X_1,X_2)) - \min_{Q\in\Delta_{P}}I_{Q}(Y:(X_1,X_2))~.
\end{align}
Note that $CoI$ refers to the co-information $CoI(Y;X_1;X_2) = I(Y:X_1)-I(Y:X_1|X_2)$ (see \cite{Bertschinger2014} for details).
For this particular choice of measures it can be shown that there is always at least one distribution $Q_0 \in \Delta_P$ for which the synergy vanishes, as was desired above. As knowledge of the pairwise marginal distributions  $P(X_1,Y)$, $P(X_2,Y)$ only specifies the problem up to any $Q \in \Delta_P$, and as the synergy varies on $\Delta_P$, we need to know the joint distribution $P(X_1,X_2,Y)$ to know about the synergy. This is indeed an intuitively plausible property and supports the functionality of the definitions given by Bertschinger and colleagues \cite{Bertschinger2014}. 

From Figure \ref{fig:PID} and the definition of $\TUNQ$, $\TSHD$, and $\TSYN$ in equations \ref{eq:TI_{unq}}-\ref{eq:TI_{syn}} it seems obvious that the following bounds hold for these measures:
 \begin{align}
     \TUNQ(Y:X_1\setminus X_2)  & \ge I_{unq}(Y:X_1\setminus X_2),  \\
     \TUNQ(Y:X_2\setminus X_1) & \ge I_{unq}(Y:X_2\setminus X_1) ,  \\
     \TSHD(Y:X_1;X_2)           & \le I_{shd}(Y:X_1;X_2),           \\
     \TSYN(Y:X_1;X_2)          & \le I_{syn}(Y:X_1;X_2) ,
 \end{align}
\noindent and this can indeed be proven, given that $I_{unq}$, $I_{shd}$, and $I_{syn}$ is taken to mean any other definition of PID that satisfies equations \ref{eq:totalMI}-\ref{eq:condMI} \cite{Bertschinger2014} and the above game theoretic assumption of a constant $I_{unq}$ on $\Delta_P$.

The measures $\TUNQ$, $\TSHD$, and $\TSYN$ require finding minima and maxima of conditional mutual information terms on $\Delta_P$. Fortunately, these constrained optimization problems are convex
for two inputs as shown in \cite{Bertschinger2014}, meaning that there is only one local minimum (maximum) which is the desired global minimum (maximum). Incorporating the constraints imposed by $\Delta_P$ into the optimization maybe non-trivial, however.
 
\subsection{PID by example: Of casinos and spies}
A short example may demonstrate the above reasoning: Let Alice and Bob bet on the outcomes $Y$ of a (perfect, etc.) Roulette table at a Casino in a faraway city, such that they do not have immediate access to these outcomes; they will only get a list of these outcomes when the Casino closes, but will have to place their bets before that. Alice has a spy $X_1$ at the Casino who informs here directly after an outcome was obtained there, but only tells the truth when the outcome was even (this includes 0). Otherwise he tells her a random possible outcome from a uniform distribution across natural numbers from 0 and 36 (just like the Roulette). Bob also has a spy $X_2$ at the casino, but in contrast to Alice's spy he only tells Bob the truth for uneven outcomes and for 0, otherwise he lies in the same way as the one of Alice, picking a random number. Neither Alice nor Bob knows about the spy of the other \footnote{This is actually irrelevant, since the game is about $Y$ only. The statement is intended for readers with a game theoretic background and should clarify that this is a trivial game, where knowledge about the opponent doesn't influence Alice's or Bob's strategy.}.
While this situation looks quite symmetric at first glance, both can prove to each other to have unique information about the outcomes at the casino, $y$. To see this, remember that Alice may suggest a game constructed by herself when trying to prove the possession of unique information. Thus, Alice could suggest to double the stakes for bets on even numbers\footnote{In roughly 50\% of the cases the outcome of the Roulette, $y$, will be even, and in these cases Alice will be told the truth. In the other 50\% of the cases, the outcome will be odd, and the spy will report a random number. Of these roughly 50\% will be even, roughly 50\% will be odd. Thus Alice will receive on average roughly 75\% even and 25\% odd numbers. Of the even numbers 2/3 will be correct. Of the odd numbers only 1/18 will be correct -- by chance. For Bob the situation is reversed. Forcing higher stakes for even outcomes will, therefore, be an advantage for Alice.}. At the end of the day, both Alice and Bob will have won a roughly equal amount of bets, but the bets Alice will typically have won payed out more, and Alice wins. In the same way, Bob could suggest to double the stakes for uneven outcomes if it were his turn to prove the possession of unique information. Thus, both have the same amount of information about the outcomes at the casino, but a part of that information is about different outcomes.

In this example, there is also redundancy as both will have the same information about the outcome $Y=0$.

 It is left for the reader to verify that Alice and Bob will gain some information (i.e. synergy) by combining what their spies tell them, but that this is not enough to be certain about the the outcome of the Roulette, i.e. $I(Y:X_1,X_2)<H(Y)$~\footnote{Hint: Think about what they can conclude if the parities of their outcomes match, and what if they don't match.}. 

\subsection{Estimating Synergy and PID for jointly Gaussian variables} 
While synergy, shared and unique information are already difficult to estimate for discrete variables, it is not immediately clear how to extend the definitions to continuous variables in general. Barrett has made significant advances in this direction though by considering PID for jointly Gaussian variables \cite{Barrett2014}. Approaches to Gaussian variables are important analytically because the classical information theoretic terms there may be computed directly from the covariance matrix of $Y$, $X_1$, $X_2$, and are important empirically due to the wide use of Gaussian models to simplify analysis (e.g. in neuroscience).

First, Barrett was able to demonstrate the existence of cases of non-zero quantities for each of synergy and shared information for such variables. This was done without reference to any specific formulation of PID measures by examining the `net synergy' (synergy minus shared information), i.e. $I(Y:X_1,X_2)-I(Y:X_1)-I(Y:X_2)$, which provides a sufficient condition for synergy where it is positive and for shared information where it is negative.
This was an important result, since the intuition of many authors was that the linear relationship between such Gaussian variables could not support synergy.

Next, Barrett demonstrated a unique form for the PID for jointly Gaussian variables which satisfies the original axioms of Williams and Beer \cite{PaulWilliamsPID} as well as having unique and shared information terms depending only on the marginal distributions $(X_1,Y)$ and $(X_2,Y)$ (as argued by Bertschinger et al. \cite{Bertschinger2014} above, and consistent with \cite{harder_bivariate_2013,Griffith2014a}).
To be specific, this unique form holds only for a \emph{univariate} output (though multivariate inputs are allowed).
This formulation maps the shared information to the \emph{minimum} of the marginal mutual information terms $I(Y: X_1)$ and $I(Y: X_2)$ -- hence is labeled the \emph{Minimum Mutual Information} (MMI) PID -- and the other PID terms follow from equations \ref{eq:totalMI}-\ref{eq:condMI}.
Interestingly, this formulation always attributes zero unique information to the input providing less information about the output.
Furthermore, synergy follows directly as the additional information provided by this ``weaker'' input after considering the ``stronger'' input. Some additional insights into this behaviour have recently been provided by Rauh and colleagues in \cite{olbrich_information_2015}
 
\section{Learning rules for maximizing $F$ and for learning the coherent infomax goal function $F_{CIP}$}
 \label{AppendixCIP}
 We here briefly present the learning rules for gradient ascent learning of neural processor learning to maximize the goal function $F$ from equation \ref{eq:F}. We only consider the basic case of a single a neural processor with binary output $Y$ here \cite{kay_contextually_1998,KayBookchapter1999}. The inputs to this processor are partitioned into two groups $\{X_{1i}\}$, representing the driving inputs and $\{X_{2j}\}$, representing the contextual inputs. These inputs enter the information theoretic goal function $F(X_1,X_2,Y)$ via their weighted sums per group as:
 \begin{align}
 \label{eq:SumDrive}
  X_1=\sum_{i=1}^m w_i X_{1i} - w_{0}=\mathbf{w}^T\mathbf{X_1}- w_{0} \\
  \label{eq:SumContext}
  X_2=\sum_{j=1}^n v_j X_{2j} - v_{0}=\mathbf{v}^T\mathbf{X_2}- v_{0}.
 \end{align}

 The inputs affect the output probability of the processor via an activation function  $A(x_1,x_2)$ as:
 \begin{equation}
 \label{eq:Act_func}
  \Theta \equiv p(Y=1 | X_1=x_1, X_2=x_2)= \frac{1}{1+\exp(-A(x_1,x_2))}.
 \end{equation}
\noindent For the sake of deriving general learning rules, $A$ may be any general, differentiable nonlinear function of the input. Note that $\Theta$ fully determines the information theoretic operation that the processor performs. $\Theta$ is a function of the weights used in the summation of the inputs. Thus, learning a specific information processing goal can only be done via learning these weights -- assuming that the input distributions of the processor can not be changed. Learning rules for these weights will now be presented.

To write the learning rules in concise form, the additional definitions:
\begin{align}
 E=\langle \Theta \rangle_{\mathbf{x_1},\mathbf{x_2}}\\
 E_{\mathbf{x_2}}=\langle \Theta \rangle_{\mathbf{x_1}|\mathbf{x_2}}\\
 E_{\mathbf{x_1}}=\langle \Theta \rangle_{\mathbf{x_2}|\mathbf{x_1}}~
\end{align}
\noindent are introduced to abbreviate the expectation of the activation across all input vectors $\mathbf{x_1}=[x_{11}\ldots x_{1m}]$, $\mathbf{x_2}=[x_{21}\ldots x_{2n}]$. These expectations are functions of the input distributions as well as of the weights and have to be recomputed after weight changes. Using online learning therefore necessitates computing these expectations over a suitable time window of past inputs. To write the learning rules in concise notation a non-linear floating average $\bar O$ of the above expectations is introduced as:

\begin{equation}
 \bar O = \Phi_0 \log \frac{E}{1-E} - (\Phi_0-\Phi_2)\log \frac{E_{\mathbf{x}_1}}{1-E_{\mathbf{x}_1}} -(\Phi_0-\Phi_1) \log \frac{E_{\mathbf{x}_2}}{1-E_{\mathbf{x}_2}}~.
\end{equation}

Using this notation, the gradients for the updates of weights $\mathbf{w}=[w_1 \ldots w_m ]$,$\mathbf{v}=[v_1 \ldots v_n ]$, and the bias coefficients $w_0$, $v_0$ are:
\begin{align}
 \frac{\partial F}{\partial \mathbf{w}}=\left\langle \left((\Phi_1+\Phi_2-\Phi_3 -\Phi_0)A-\bar O \right)\frac{\partial A}{\partial x_1}\Theta(1-\Theta)\mathbf{x_1} \right\rangle_{\mathbf{x_1},\mathbf{x_2}}\\
 \frac{\partial F}{\partial \mathbf{v}}=\left\langle \left((\Phi_1+\Phi_2-\Phi_3 - \Phi_0)A-\bar O \right)\frac{\partial A}{\partial x_2}\Theta(1-\Theta)\mathbf{x_2} \right\rangle_{\mathbf{x_1},\mathbf{x_2}}\\
 \frac{\partial F}{\partial w_0}=\left\langle \left((\Phi_1+\Phi_2-\Phi_3 -\Phi_0)A-\bar O \right)\frac{\partial A}{\partial x_2}\Theta(1-\Theta)(-1) \right\rangle_{\mathbf{x_1},\mathbf{x_2}}\\
 \frac{\partial F}{\partial v_0}=\left\langle \left((\Phi_1+\Phi_2-\Phi_3 -\Phi_0)A-\bar O \right)\frac{\partial A}{\partial x_2}\Theta(1-\Theta)(-1) \right\rangle_{\mathbf{x_1},\mathbf{x_2}}
\end{align}

Last, we note that for the specific implementations of CIP, the activation function was chosen as:
\begin{equation}
 A(x_1,x_2)= x_1 \left[k_1 +(1-k_1)\exp(k_2 x_1 x_2) \right]~,
\end{equation}
\noindent with $0\leq k_1<1$ and $k_2 >0$, and with $x_1$, $x_2$ being realizations of $X_1$, $X_2$ from equations \ref{eq:SumDrive}, \ref{eq:SumContext}.  This specific activation function \cite{KayBookchapter1999,phillips_discovery_1995} guarantees that:
\begin{itemize}
 \item Zero output activation can only be obtained if the summed driving input $X_1$ is zero.
 \item For zero summed contextual input $X_2$, the output equals the summed driving input.
 \item A summed contextual input of the same sign as the summed driving input leads to an amplification of the output. The reverse holds for unequal signs.
 \item The sign of the output is equal to the sign of the summed driving input.
\end{itemize}
\noindent These four properties were seen as essential for an activation function that supports coherent infomax.
\end{appendix}


\end{document}